\documentclass[12pt]{iopart}
\usepackage{graphicx,color}
\usepackage{iopams}

\begin{document}

\paper{Stability of Small Neutral and Charged Strontium Clusters}

\author{A Lyalin\dag, 
        A V Solov'yov\ddag\footnote[3]{On leave from A. F. Ioffe Physical-Technical Institute, 
                                       194021 St. Petersburg, Russia}, 
        C Br\'{e}chignac$\|$ 
        and W Greiner\ddag  }

\address{\dag\, Institute of Physics, St Petersburg State University, \\
Ulianovskaya str. 1, 198504 St Petersburg, Petrodvorez, Russia}

\address{\ddag\, Frankfurt Institut for Advanced Studies, Johann Wolfgang Goethe University,
Robert-Mayer Str. 10, D-60054 Frankfurt am Main, Germany}

\address{$\|$ Laboratoire Aim\'{e} Cotton CNRS II, Universit\'{e} Paris-Sud, \\
F-91405 Orsay CEDEX, France}

\eads{\mailto{lyalin@rpro.ioffe.rssi.ru}, \mailto{solovyov@fias.uni-frankfurt.de}}

\begin{abstract}
Dissociation and fission of small neutral, singly and doubly charged 
strontium clusters are studied 
by means of {\it ab initio} density functional theory methods 
and high-resolution time-of-flight mass spectrometry.
Magic numbers for small strontium clusters possessing enhanced stability towards 
monomer evaporation and fission are determined. It is shown that ionization 
of small strontium clusters results in the alteration of the magic numbers.
Thermal promotion of the Coulombic fission for the $Sr_7^{2+}$ cluster is predicted.
\end{abstract}

\maketitle

Processes leading to the instability and the fission 
of metal clusters are among the most fundamental in cluster science,
see e.g. \cite{ISACC-book04}. Investigation of 
the metal clusters decay provides a direct tool for 
studying  intrinsic stability and binding forces of these objects.
Such investigations attract an increased interest,
because the general features of metal clusters decay and nuclear fission
are quite similar. 

One can distinguish two classes of phenomena in the decay of metal clusters. 
The first one is dissociation or evaporation of fragments due 
to vibrational excitation, that makes cluster thermally metastable. 
The second one is fission that occurs when the repulsive Coulombic 
forces associated with an excess of charge overcome the electronic 
binding energy of the cluster \cite{Sattler81,Naher97,Heinebrodt97,Yannouleas99,HFLDA1,HFLDA2}.
Multiply charged metal clusters are stable towards fission when 
their size exceeds the critical size of stability, which depends
on the type of metal species and the cluster charge 
\cite{Martin84,Brechignac89}.
In the latter case evaporation is the dominant channel of cluster decay. 
For the cluster sizes smaller than the critical size fission becomes more favorable 
\cite{Brechignac90}.  

The most sophisticated situation arises when the cluster size 
approaches the ``critical size" region. It has been found experimentally that in this case 
the internal thermal excitation can influence fission channels and promote the Coulombic 
fission \cite{Brechignac98}. On the other hand fission into two charged fragments 
can stimulate an additional ejection of neutral atom during or immediately after the system 
overcomes the fission barrier. Such an interplay between the Coulombic fission and the evaporation
processes has recently been observed \cite{Brechignac04}. 
 
In this Letter we report the results of theoretical and experimental investigation 
of stability of small neutral, singly and doubly charged strontium clusters 
towards the emission of neutral and singly charged fragments. 
We show that the closure of electronic shells of the valence electrons 
enhances the stability of small strontium clusters towards monomer evaporation. 
We demonstrate that the ionization of small strontium clusters results 
in the alteration of the magic numbers for strontium clusters. 
By {\it ab initio} molecular dynamics simulations we determine the critical 
appearance size for doubly charged strontium clusters as well as the region 
of cluster sizes in which the strong competition between evaporation and 
fission takes place. We predict theoretically the thermal promotion of the 
Coulombic fission for the $Sr_7^{2+}$ cluster and confirm the assumption made in  
\cite{Brechignac04} about the strong shape deformation of 
the fissioning $Sr_7^{2+}$ cluster.

The experimental setup is similar to that used in previous 
works \cite{Brechignac04,Brechignac_in_Martin_96}. 
Calculations have been performed with the use of  a core-polarization potential
to simulate the $1s^2 2s^2 2p^6 3s^2 3p^6 3d^{10}$ core electrons of the Sr atom. 
The density-functional theory based on the hybrid Becke-type three-parameter 
exchange functional 
paired with the gradient-corrected Perdew-Wang 91 correlation functional (B3PW91) 
has been used throughout this work (see, e.g., \cite{Chemistry,Dobson-DFT} 
and references therein). 
Such an approach has proved to be a reliable 
tool for {\it ab initio} studies of the structure and properties 
of strontium clusters \cite{Wang01}.
To find the optimized geometry of a cluster we have used the procedure 
described in \cite{struct_Na,struct_Mg}.
The results of cluster geometry optimization and analysis of  
various isomer forms for neutral, singly and doubly charged
small strontium clusters will be discussed in detail elsewhere \cite{struct_Sr}.
Calculations have been carried out with the use of
the GAUSSIAN 03 software package \cite{Gaussian}.
The SDD basis set 
of primitive Gaussians has been used to expand the cluster 
orbitals (see, e.g., \cite{Chemistry} and references therein).

 \begin{figure}[h]
\begin{center}
\includegraphics[scale=1.,clip]{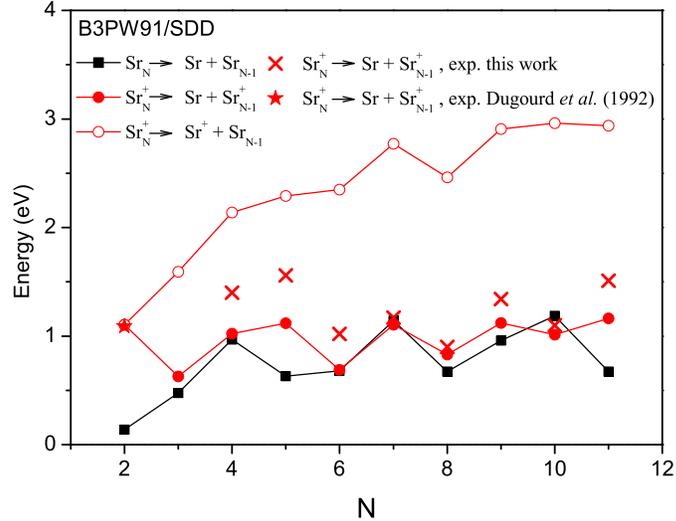}
\end{center}
\caption{Monomer dissociation energies for neutral and singly charged $Sr$-clusters. 
Filled squares and circles:  neutral monomer dissociation energies, 
$D_{N,1}^{0,0}$, and, $D_{N,1}^{+,0}$, for 
neutral and singly charged $Sr$-clusters, respectively. 
Open circles: singly charged monomer dissociation energies, $D_{N,1}^{+,+}$, 
for singly charged $Sr$-clusters. 
Crosses and star: experimental data.}
\label{dissociation_0_P}
\end{figure}

Figure \ref{dissociation_0_P} shows the monomer dissociation energies 
for neutral and singly charged $Sr$-clusters as a function of the cluster size.
Filled squares and circles represent neutral monomer dissociation energies, 
$D_{N,1}^{Z+,0} = E_{tot}(M_{1}^{}) + E_{tot}(M_{N-1}^{Z+}) - E_{tot}(M_{N}^{Z+})$,
calculated for neutral and singly charged strontium clusters respectively. 
Here $E_{tot}(M_{N}^{Z+})$ is the total energy of an optimized $N$-particle metal 
cluster with charge $Z+$.
Crosses in figure \ref{dissociation_0_P} show the experimental results 
for neutral monomer dissociation energies $D_{N,1}^{+,0}$ 
of singly charged strontium clusters obtained in this work. 
Star presents the experimental value of
$D_{2,1}^{+,0}$ for the ionized strontium dimer from reference \cite{Dugourd92}. 

The local maxima in the size dependence of the monomer dissociation energy, 
$D_{N,1}^{0,0}$, at $N=$ 4, 7 and 10 correspond to the most stable configurations of  
neutral $Sr$-clusters towards monomer evaporation $Sr_{N} \rightarrow Sr + Sr_{N-1}$. 
The same magic numbers have also been obtained from the analysis 
of binding energies of small neutral strontium \cite{struct_Sr} 
and magnesium \cite{struct_Mg} clusters.
The electronic configuration of the strontium atom is $[Kr]5s^2$, which means that there 
are two valence electrons per atom. 
Accounting for the semi-core $4p$ electrons of strontium simultaneously with the 
valence electrons increases the absolute value of the binding energy by about 10-20\%
although it does not change the general qualitative trends in the properties 
of small strontium clusters \cite{Kumar01}.
Therefore, one can state that
the most stable magic clusters $Sr_4$, $Sr_7$ and $Sr_{10}$  
possess $N_{el}=$ 8, 14 and 20 valence electrons respectively, which  
is in agreement with the deformed jellium model (see, e.g., \cite{MLSSG03,93,66,62} 
and references therein as well as discussion in \cite{struct_Mg}). 

Open  circles in figure \ref{dissociation_0_P} 
show the singly charged monomer dissociation energies,  
$D_{N,1}^{Z+,+} = E_{tot}(M_{1}^{+}) + E_{tot}(M_{N-1}^{(Z-1)+}) - E_{tot}(M_{N}^{Z+})$,
calculated for singly charged  strontium clusters.
Figure \ref{dissociation_0_P} demonstrates that the evaporation of a charged monomer 
for singly charged  $Sr$-clusters,
$Sr_{N}^{+} \rightarrow Sr^{+} + Sr_{N-1}$, is strongly suppressed in comparison
with the evaporation of a neutral strontium atom,
$Sr_{N}^{+} \rightarrow Sr + Sr_{N-1}^{+}$.  
The singly charged strontium dimer $Sr_{2}^{+}$ is more stable towards decay 
in comparison with the neutral dimer. This phenomenon has a simple physical explanation: 
the removed electron is taken from the antibonding orbital, and thus cationic
strontium dimer is stronger bounded. The similar effect has been 
discussed for cationic magnesium clusters in \cite{struct_Mg}.

The local maxima in the size dependence of the neutral monomer dissociation energy,
$D_{N,1}^{+,0}$, for the $Sr_{5}^{+}$ and $Sr_{7}^{+}$ clusters indicate 
their enhanced stability towards monomer evaporation.
Figure \ref{dissociation_0_P} clearly demonstrates that the single ionization of small 
strontium clusters results in the alteration of the magic numbers. 
The similar alteration of the magic number from 
$N =$ 4 for neutral to $N =$ 5 for cationic magnesium clusters has 
been noticed in our recent work \cite{struct_Mg}. This fact can be explained 
by the manifestation of shell effects. The singly charged alkaline earth metal clusters
always possess odd number of valence electrons and, thus, always contain open 
electronic shells.  In this case the enhanced stability of a singly charged 
alkaline earth metal cluster ion arises, when 
the electronic configuration of the ion has one hole in or an extra electron 
above the filled shells \cite{struct_Mg}. Thus the electronic configuration 
containing an extra electron becomes more favorable for $Sr_{5}^{+}$.

The calculated values of $D_{N,1}^{+,0}$ are in a good qualitative 
agreement with the experimental results. Theoretical curve reproduces all 
the features in the size dependence of the dissociation energy obtained 
in experiment. However,  
the calculated monomer dissociation energies for the $Sr_4^{+}$, $Sr_5^{+}$ and $Sr_6^{+}$ clusters
underestimate the 
experimental values  by approximately 0.4 eV.  
This discrepancy can be  attributed to 
the distribution of daughter fragments registered in experiment
over different isomer states.
One can assume that after evaporation of a neutral monomer 
the resulting daughter fragment 
remains not always  in the ground state, but in one of the higher energy isomer states. 
This brings the experimentally measured dissociation energies up. 
This fact is not taken into account in the calculations reported.
Therefore the monomer dissociation energies calculated theoretically 
lay lower those measured in the experiment.  
The results of geometry optimization 
of different isomer states of small strontium clusters 
will be discussed elsewhere  \cite{struct_Sr}.

\begin{figure}[h]
\begin{center}
\includegraphics[scale=1.,clip]{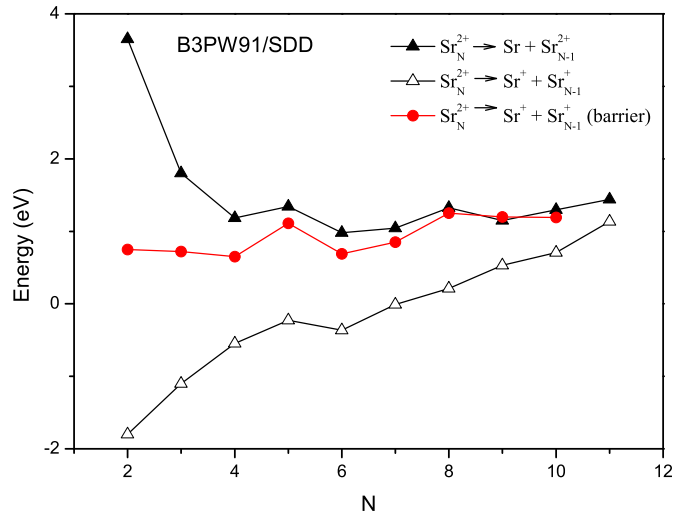}
\end{center}
\caption{Monomer dissociation energies for doubly charged $Sr$-clusters.
Filled and open triangles: neutral and singly charged monomer 
dissociation energies, $D_{N,1}^{2+,0}$ and $D_{N,1}^{2+,+}$ respectively.
Filled circles: fission barriers for the process 
$Sr_{N}^{2+} \rightarrow Sr^{+} + Sr_{N-1}^{+}$.
}
\label{dissociation_PP}
\end{figure}

Figure \ref{dissociation_PP} shows the monomer dissociation energies 
for doubly charged $Sr$-clusters as a function of cluster size.
Filled and open triangles represent the neutral and singly charged monomer 
dissociation energies, $D_{N,1}^{2+,0}$ and $D_{N,1}^{2+,+}$, respectively.
It is important to note that the decay of doubly charged parent cluster 
into two singly charged daughter fragments usually involves the overcoming 
a fission barrier caused by the Coulombic repulsion, 
while evaporation is the barrierless process \cite{Yannouleas99,Brechignac90}.  
Therefore, the values $D_{N,1}^{2+,+}$
characterize the energy balance 
between the initial and the final states of the system and do not 
give direct information about the fragmentation rate,
because the fission rate is usually determined  by the fission barrier. 
Filled circles in figure  \ref{dissociation_PP} show the size dependence
of the fission barriers calculated for the 
process $Sr_{N}^{2+} \rightarrow Sr^{+} + Sr_{N-1}^{+}$.
To calculate the fission barriers we use the procedure described in detail in our 
recent works \cite{Lyalin04,Lyalin04a,Lyalin_ISACC04}.
The doubly charged strontium clusters with the number of atoms $N=$ 5 and 8 
possess enhanced stability towards the neutral monomer evaporation as well as towards 
the Coulombic fission with ejection of a singly charged monomer.

For small doubly charged strontium clusters with the number of atoms $N \le 7$ 
the neutral monomer dissociation energy, $D_{N,1}^{2+,0}$, 
exceeds significantly the fission barrier. Therefore, in this cluster size region fission 
prevails over neutral monomer evaporation. As the cluster size increases 
the fission barrier becomes comparable and exceeds the dissociation energy.
In the size region $8 \le N \le 10$ the height of the fission barrier 
and $D_{N,1}^{2+,0}$ become almost equal resulting in the competition 
of the fission and the evaporation processes. 
For larger cluster sizes neutral monomer evaporation dominates over fission.

\begin{figure}[h]
\begin{center}
\includegraphics[scale=1.,clip]{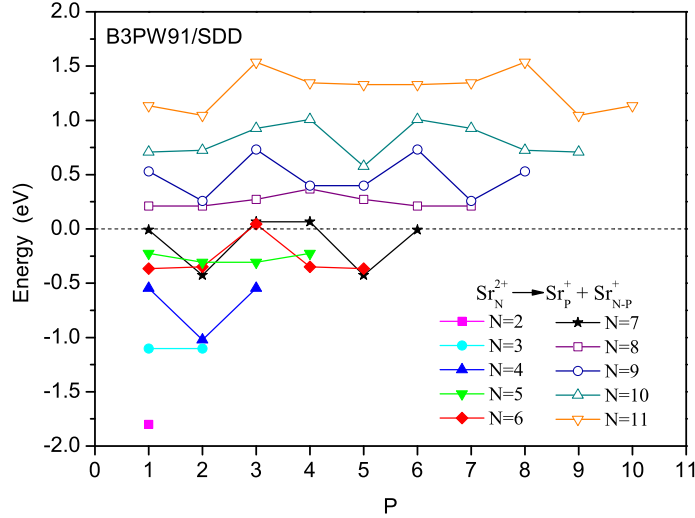}
\end{center}
\caption{Energy release $D_{N,P}^{2+,+}$ for different fission channels $P$ in the process
$Sr_N^{2+} \rightarrow Sr_P^{+} + Sr_{N-P}^{+}$.}
\label{dissociation2}
\end{figure}

To find the critical size of stability at which clusters undergo Coulombic fission
it is necessary to analyze
the energy balance for all possible fission channels as a 
function of cluster size. 
Figure \ref{dissociation2} shows the energy release, $D_{N,P}^{2+,+}$, as a function of 
different fission channels $P$ in the process $Sr_N^{2+} \rightarrow Sr_P^{+} + Sr_{N-P}^{+}$
for the doubly charged strontium clusters with the number of atoms up to $N=$ 11.
Coulombic fission takes place when the energy release is negative, which means 
that the final state of the system is energetically more favorable 
in comparison with the initial state of the parent cluster.  
We found that the critical {\it appearance} size for the 
doubly charged strontium clusters is equal to $N_{app} = $ 8.
For the strontium clusters with $N \ge 8$ the energy release is positive for all   
fission channels. In this case cluster can decay via fission only if it 
possesses enough vibrational energy to promote the Coulombic fission.

The results of our {\it ab initio} calculations of the critical appearance size for 
doubly charged strontium clusters are in a good agreement with those derived from 
experiment \cite{Heinebrodt97,Brechignac98}. Doubly charged strontium clusters 
with number of atoms $N \ge N_{app}$ are directly observed in the mass spectrum.

Figure \ref{dissociation2} demonstrates strong influence of the shell effects on the 
fragmentation process $Sr_N^{2+} \rightarrow Sr_P^{+} + Sr_{N-P}^{+}$. 
Thus,  ejection of the singly charged dimer $Sr_2^{+}$ is the 
energetically favorable channel of decay
for the $Sr_4^{2+}$, $Sr_7^{2+}$, $Sr_9^{2+}$ and
$Sr_{11}^{2+}$ clusters.  
For the $Sr_{10}^{2+}$ cluster, the symmetric  
fission channel $Sr_{10}^{2+} \rightarrow 2 Sr_5^{+} $ is more favorable energetically.
Singly charged strontium clusters always possess odd number of valence electrons
and, thus, the interpretation of shell effects in terms of electronic shell 
closings is not  straightforward. 
In this case, the enhanced stability of 
singly charged strontium clusters arises, when the electronic configuration 
of the cluster has one hole 
in or an extra electron above the filled shells \cite{struct_Mg}. 
This rule explains the manifestation of
shell effects in fission of doubly charged strontium clusters.

\begin{figure}[t]
\begin{center}
\includegraphics[scale=1.,clip]{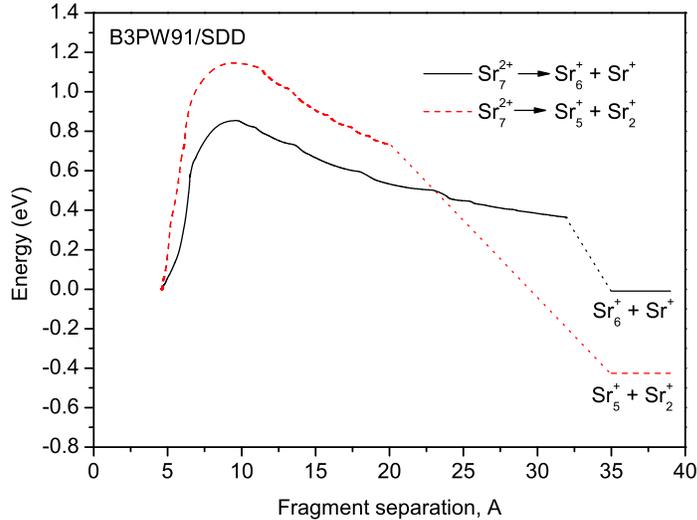}
\end{center}
\caption{Fission barriers for the $Sr_7^{2+}$ cluster
as a function
of distance between the centers of mass of the fragments. }
\label{barrier_7++}
\end{figure}

The most complex situation with fission occurs when the cluster size lays   
in the ``critical size" region. Thus, the $Sr_7^{2+}$ cluster is the largest
doubly charged strontium cluster which 
can spontaneously decay via the Coulombic fission. The following two channels 
$Sr_7^{2+} \rightarrow Sr_6^{+} + Sr^{+}$  and  
$Sr_7^{2+} \rightarrow Sr_5^{+} + Sr_2^{+}$ are allowed energetically.
Fission via the ejection of the dimer  $Sr_2^{+}$ is much more favorable from 
the energetic view point, however, in this case the system must 
overcome the higher fission barrier, as it is seen 
from figure \ref{barrier_7++}.  
Therefore  the ejection of the singly charged monomer 
$Sr^{+}$ is the dominant fission channel for low cluster temperatures.
As the temperature increases, the probability of ejection 
of the dimer $Sr_2^{+}$ grows. The similar effect was observed 
for the triply-charged strontium cluster $Sr_{19}^{3+}$ 
in \cite{Brechignac98}.

The fission barrier maxima are located at small separation 
distances 
($d_{max} \approx 9.5$ \AA) 
just before the scission point. 
At such distances the parent cluster is strongly deformed. 
This shape deformation influences the dynamics of the fission and can induce, 
by dissipative effects, the ejection of a fast neutral atom \cite{Brechignac04}.

In conclusion, experimentally measured dissociation energies are in  
agreement with those derived from our {\it ab initio} calculations.
We determine the cluster magic numbers  possessing the enhanced stability towards 
monomer evaporation and fission. We demonstrate that the ionization of small 
strontium clusters results in the alteration of the magic numbers. 
The critical appearance size  for the doubly charged strontium 
clusters determined  theoretically is in a good agreement with experimental observations.
Thermal promotion of the Coulombic fission for 
the $Sr_7^{2+}$ cluster is predicted.

\ack
The authors acknowledge support of this work by  INTAS {(grant No 03-51-6170)} and
the Russian Foundation for Basic Research {(grant No 03-02-16415)}.
We gratefully acknowledge support by the Frankfurt Center for Scientific Computing. 
A.L. expresses his gratitude to the Alexander von Humboldt Foundation
for financial support.

\end{document}